\documentclass[twocolumn,apjl]{emulateapj}

\def\plotfiddle#1#2#3#4#5#6#7{\centering \leavevmode
    \vbox to#2{\rule{0pt}{#2}}
    \includegraphics{#1}}

\begin{document}

\title{Unveiling the Boxy Bulge and Bar of the Andromeda Spiral Galaxy}

\author{
Rachael L.\ Beaton\altaffilmark{1},
Steven R.\ Majewski\altaffilmark{1},
Puragra Guhathakurta\altaffilmark{2},
Michael F.\ Skrutskie\altaffilmark{1},\\
Roc M. Cutri\altaffilmark{3}, 
John Good\altaffilmark{3},
Richard J.\ Patterson\altaffilmark{1}, 
E.\ Athanassoula\altaffilmark{4},
and Martin Bureau\altaffilmark{4}}

\altaffiltext{1}{Department of Astronomy, University of Virginia,
Charlottesville, Virginia 22903-0818; rlb9n@virginia.edu, srm4n@virginia.edu,
mfs4n@virginia.edu, rjp0i@virginia.edu}
\altaffiltext{2}{UCO/Lick Observatory, Department of Astronomy \&
Astrophysics, University of California, Santa Cruz, California 95064;
raja@ucolick.org}
\altaffiltext{3}{Infrared Processing and Analysis Center,  Caltech, MS 100-22,
Pasadena, CA 91125; roc@ipac.caltech.edu, jeg@ipac.caltech.edu}
\altaffiltext{4}{Observatoire de Marseille, 2 place Le Verrier, 13248 Marseille
Cedex 04, France; lia@oamp.fr}
\altaffiltext{5}{Sub-Department of Astrophysics,University of Oxford,
Denys Wilkinson Building, Keble Road, Oxford OX1 3RH, United Kingdom; 
bureau@astro.ox.ac.uk}

\begin{abstract}

A new, 2.8 deg$^2$  $J,H,K_s$ infrared survey from the 2MASS 6x 
program across the extent of the optical disk of the Andromeda 
(M31) galaxy provides a clear view of the M31 center almost 
completely unfettered by dust extinction, and reveals a high 
contrast bulge with extremely boxy isophotes dominating the NIR
light to a semi-major axis of $\sim$$700$$''$ (2.6 kpc).  
The inner bulge ($\lesssim50$$''$) is relatively circular, 
but shows some isophotal twisting.  Beyond this, the ellipticity 
and boxiness of the bulge increase with radius --- achieving a 
boxiness that rivals that of any other known disk galaxy observed in the 
near infrared --- and the position angle is constant at 
$\sim$$50^{\circ}$, which is about 10$^{\circ}$ higher than the 
position angle of the M31 disk.  
Boxy bulges in highly inclined disks have been shown to be the 
vertical structure of bars, and self-consistent, $N$-body 
modeling specific to the NIR images presented here can reproduce 
the observed NIR M31 features with a combination of a classical 
bulge and a boxy bulge/bar.  
Beyond the boxy bulge 
region and nearly along the 40$^{\circ}$ position angle of the 
disk a narrow ridge of infrared flux, which can be identified 
with the thin part of the bar, more or less
symmetrically extends into the inner disk at semi-major axis 
radii of 700$''$ to 1200$''$ or more.  
Little variation in the 
morphology or relative brightnesses of these various 
M31 structures is 
seen across the NIR bands (i.e., no color gradients are seen).  
These new data verify that M31 is a barred spiral galaxy 
like the Milky Way.

%The observed ansae provide direct evidence that the M31 bar 
%extends beyond its boxy bulge.

\end{abstract}


\section{Introduction}

For long after the discovery that spiral nebulae are ``island universes'', the Andromeda
Galaxy (M31) was thought to present a reasonable mirror image of our own Milky
Way.  This morphological mirror, however, was clouded with the discovery of barred structure in the Milky Way,
revealed both as a gas dynamical structure (\citealt{blitz93}) and a stellar feature (Weinberg 1992).
More recent evidence (e.g., Parker et al. 2003, 2004; see also 
\citealt{newberg04}) suggests that the Milky Way's central bar
may exert significant influence, or at least account for observable stellar density asymmetries, 
to large distances from the Galactic mid-plane.
%
% Something about the dimensions in kpc.
%

Although there have been isolated lines of evidence supporting the
existence of a bar in M31, 
its existence has remained uncertain.  Fifty years ago
\citeauthor{Lind56} (\citeyear{Lind56}) first recognized a general
misalignment of the innermost isophotes of M31 with the major axis of the
outer isophotes.  This was interpreted as an indication that the bulge was
triaxial due to the presence of a nuclear bar. 
Stark (1977) attempted to match the observed isophotal 
twists and overall optical surface brightness distribution within 2 kpc
with a one-parameter family of triaxial models, and also concluded that
misalignment between the bulge and disk isophotes required a central bar.
This notion was supported by evidence from gas dynamics (\citealt{Stark94}, 
\citealt{Ber01}; \citealt{Ber02}).  
Kinematical work in the nucleus of M31 (\citealt{kor88}; \citealt{dress88})
revealed large rotation velocities, 
%dispersions and velocity gradients at small scales.  Though this was 
initially interpreted as evidence of a
supermassive black hole at the center of M31, but later explained using bar-like potentials
(\citealt{ger88}).
%
%\citeauthor{ger88}
%(\citeyear{ger88}) illustrated that these data could be also explained using
%bar-like potentials as supported by other observational evidence at larger radii (most
%specifically the CCD observations of \citealt{kent83}).
%
%
%This interpretation, however, is largely
%dependent on the classification by Kormendy (\citeyear{kor88}) of the nucleus
%as a disk.  Recent modeling has presented nuclear disk models that can match
%observed kinematics around the central black hole (\citealt{salow04}), adding
%further strength to the bar-like nature of M31.
%
Though Kent's (\citeyear{kent83}) optical imaging did not identify
the previously suggested misalignment of the bulge and disk isophotes, the data showed 
some degree of boxiness to the bulge isophotes, revealed by a non-zero
fourth-order harmonic in observed isophotal shapes.  
%Kent (\citeyear{kent83}) also reported an
%inability to fit surface brightness profiles with the \citeauthor{king66}
%(\citeyear{king66}) model due to a failure of the observed profile to flatten towards
%the center.  
The boxy classification of M31's bulge by \citeauthor{kent83}
(\citeyear{kent83}) makes a compelling implication of a more complicated central M31.
% may more complicated than intially prescribed. These observations, however, 
However, all previous optical studies of M31 have been substantially hindered by 
% were largely affected by 
dust contamination in this highly inclined ($i=77.5^\circ$) system.

\begin{figure*}
\plotfiddle{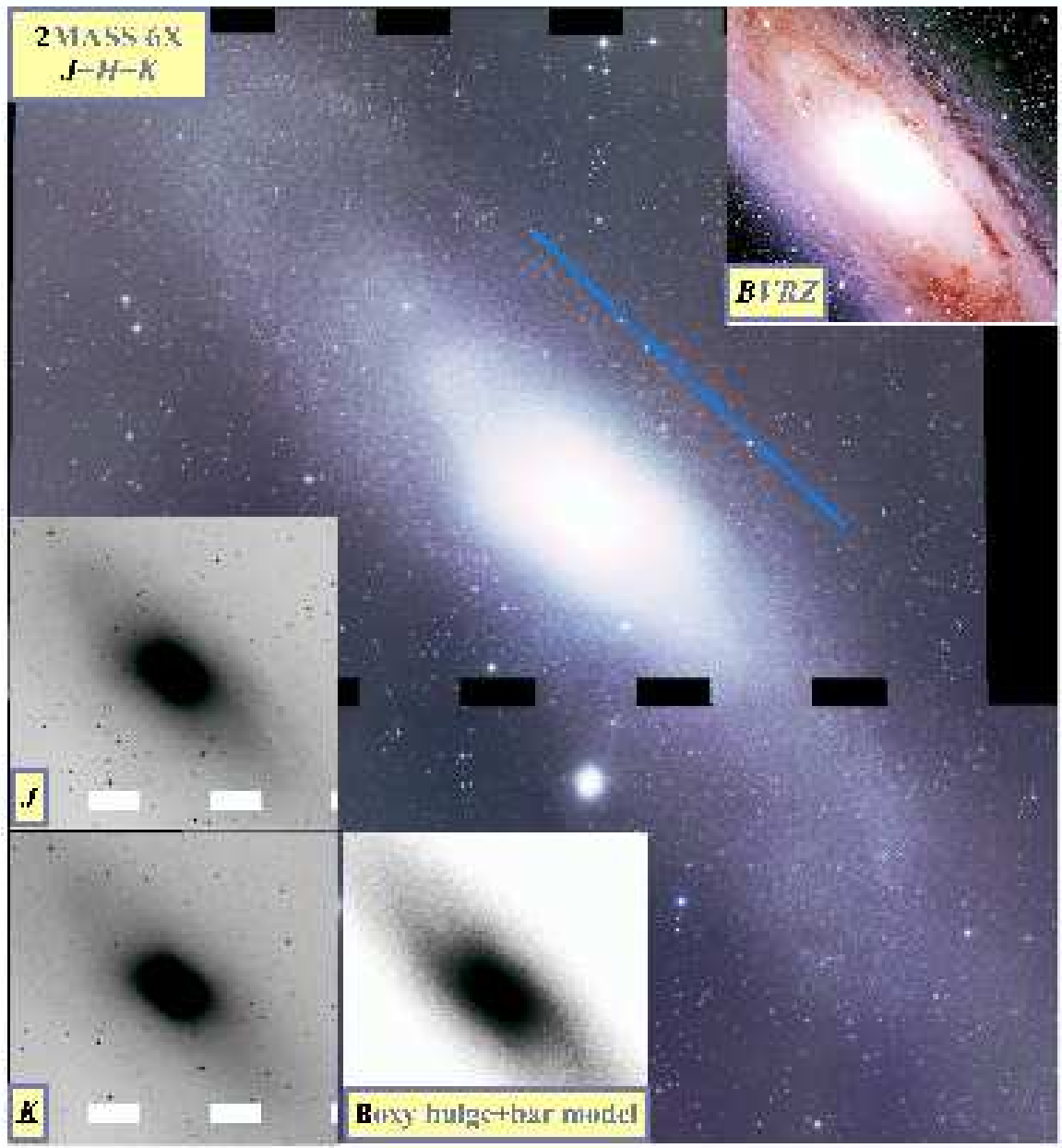}{6.6in}{0}{80}{80}{-250}{-15}

\caption{The central image is a color composite using the $J,H,K_s$ images of M31 from the 2MASS 6x observations.  The line is 
40$'$ (8kpc) long.  Two of these contributing images ($J$ and $K_s$) are shown in the lower left insets.  The upper
right inset is a color composite from the optical $B,V,R,z$ bands.  An additional inset shows a boxy bulge
plus bar model matching the system from Athanassoula \& Beaton (2006).  Insets are 30\% smaller than the
central image.
}
\end{figure*}

It has long been known that observations in the near-infrared (NIR), where the effects of dust are 
mitigated, can reveal central galaxy structures hidden at shorter wavelengths,
including multiple nuclei, bars, and boxy/peanut-shaped bulges 
(e.g., Hackwell \& Schweizer 1983, Scoville et al.\ 1988, Telesco et al.\ 1991, Majewski et al.\ 1993,
Quillen et al.\ 1997).  The central regions of
highly inclined systems are particularly challenging, but especially valuable for 
understanding the frequency of boxy/peanut morphologies.
Despite several surveys of highly-inclined 
%edge-on galaxies have recently been undertaken by 
galaxies (e.g., Shaw 1993, \citealt{lut2}, Bureau et al. 2006), perhaps
less than 100 have been studied in the NIR.  
This includes the Milky Way, for which a clear portrait of its 
central, bar-like structure has been revealed 
by the distribution of carbon stars in the
Two Micron All Sky Survey (2MASS) point source catalog (\citealt{skrut01},
Cole \& Weinberg 2002).  
Comparable studies of M31 have not been previously available because of the combined need for 
greater depth over a large field of view.  Such a view is now provided by 
deep scans of M31 using the 2MASS North facility.

\section{The 6x Observations of M31}

%The photometry for our NIR analysis of M31 comes from the northern 
%2MASS facility at Mt. Hopkins.

Near the end of the normal 2MASS observing period, long exposure observations 
were taken of a number of special target regions, including M31,
during idle times in the regular operation of the survey.  These observations, taken
with exposures six times longer than the nominal 2MASS observations and 
referred to as the ``6x observations", 
%were designed to go approximately 1 magnitude deeper than the original 2MASS 
%observations.  The 6x utilized 
used the same freeze-frame scanning as the main survey.\footnote{See
 http://www.ipac.caltech.edu/2mass/releases/allsky/doc/ for a
description of the 2MASS 6x survey.} 
M31 was observed using 1$^{\circ}$ long scans similar to those used for 2MASS
calibration field observations.  
A telescope command error produced coverage gaps of a few arcmin$^2$ in the 
2.8 deg$^2$ M31 6x mosaic.  These gaps (visible in Fig.\ 1) 
do not affect the work presented here.

We adopt the standard reduction of the 6x M31 data provided
by the 2MASS Project, which used a slightly modified verision of
the automated data reduction pipeline utilized to process the
original survey data.  The final output 6x Atlas Images are identical
in format to those from the original survey.  Composite $J$, $H$ and $K_s$ mosaics
were constructed from the individual 6x M31 Atlas Images using
the Montage\footnote{See http://montage.ipac.caltech.edu/.}
software package.  Montage matches the background
levels of all images by applying an additive offset and gradient correction
that is computed using a least-squares fit to the mean intensity offsets
of all pixels in the overlapping regions between each pair of images.
No correction was applied to the images to match the photometric zero points
between adjacent 6x scans.  However, the relative zero points variation
is $<1\%$ in all bands over the region of the M31 bulge, so there should be
no discernable effect on the surface brightness fits. 
From these mosaics, a mean ``sky" level was determined from the peak of the 
histogram of pixel values in the outer image, beyond the apparent disk of M31, 
which is defined by the ring of HII regions readily apparent in the GALEX 
images of M31 (e.g., see Thilker et al.\ 2005).  This offset value 
determined for each $J$, $H$ and $K_s$ image was subtracted to adjust pixel values to
an approximately linear representation of M31 flux; this is sufficient for our present 
goal to understand the overall morphology of the M31 center.

\begin{figure}
%\label{fig:contours}
\plotfiddle{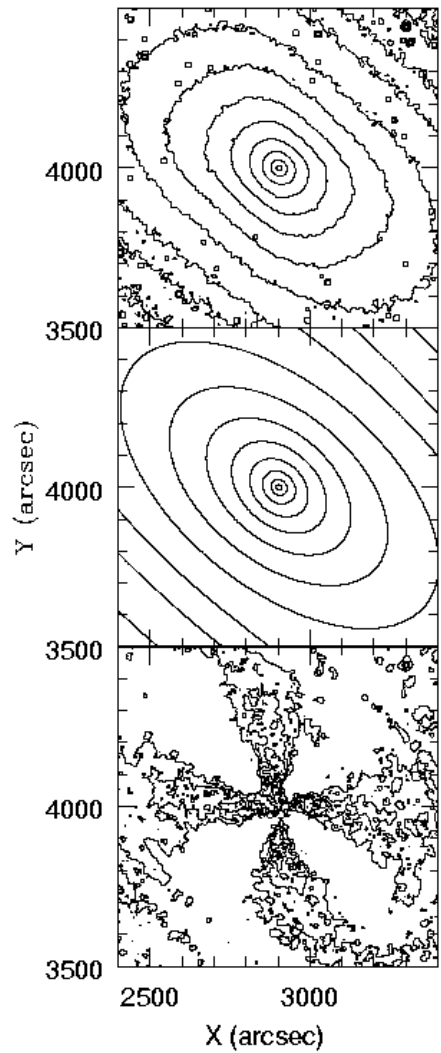}{5.3in}{0}{75}{75}{-110}{-20}
%\plotone{plotcontsml.eps}
\caption{{\it Top:} Contour plot for the central $K_s$ band image.  {\it Middle:}
Best fit ellipses to the $K_s$ data.  {\it Bottom:} Residual plots created by subtracting the best fitting 
ellipse from the $K_s$ data.  The contours in all three panels have uniform logarithmic spacing.
The pixel coordinates are pixel values in the 6x image; the image scale is  $1''$\,pixel$^{-1}$.
The $X$ and $Y$ coordinates are aligned with the right ascension and declination axes.} 
\end{figure}

\begin{figure}
%\epsscale{1.00}
\plotfiddle{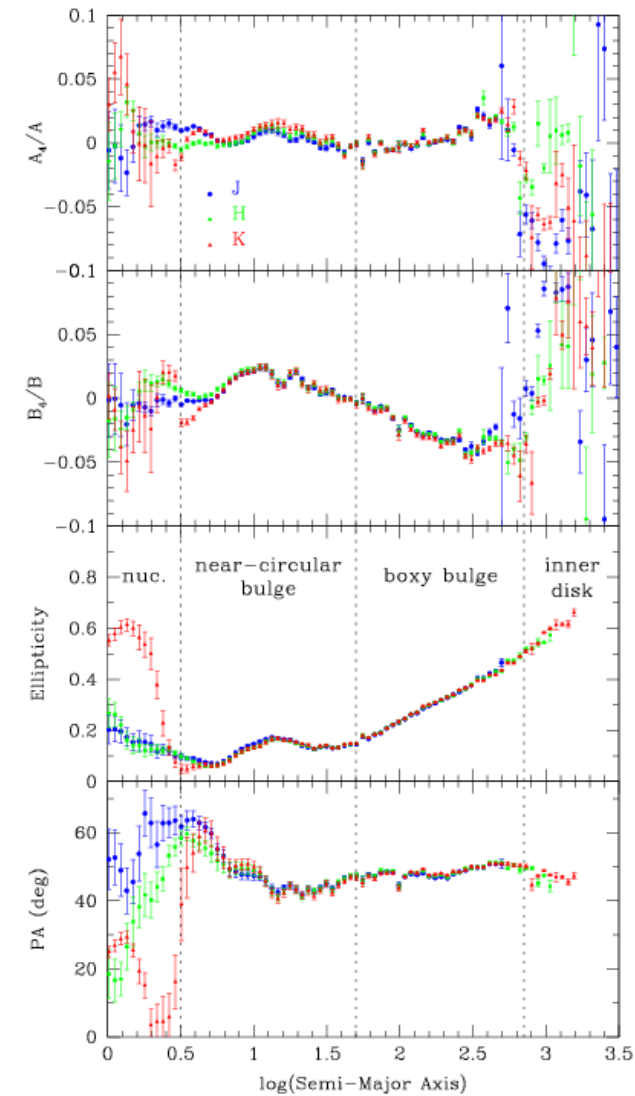}{5.5in}{0}{80}{80}{-145}{-28} 
\caption{$A_4/A$, $B_4/B$, ellipticity and position angle of M31 as a function 
of semi-major axis (in arcseconds) as measured in the $J$, $H$ and $K_s$ bands (blue, green and red points,
respectively).}
\end{figure}

\section{The Near-Infrared Structure of M31 }

A boxy, high surface brightness central bulge is the most obvious M31 feature 
in the 6x images, and this significantly distinguishes M31's NIR morphology 
from that seen in the optical (Fig.\ 1). 
Apart from slightly more dust modulation in $J$, the basic
character of M31 system is the same across the NIR bands, most evidently 
in the similarity of the high contrast region of the bulge with its 
$\sim$$700$$''$ major axis radius
and, at fainter surface brightness, the appearance of a thin ridge of
NIR light just beyond this bulge nearly along the disk major axis.

To quantify this morphology, ellipses were fit to the isophotal contours 
in each the $J, H$ and $K_s$ bands using the ELLIPSE package in IRAF (Fig.\ 2). 
This code uses the techniques of \citeauthor{jed87}  (\citeyear{jed87}) to fit  
a Fourier series to concentric galactic isophotes. The fourth order 
coefficients ($A_4$ and $B_4$) of the best fit Fourier series are a useful descriptor 
of the deviation of the isophotes from true ellipses, from ``boxy" ($B_4$$<$0, $A_4$$>$0)
to ``disky" ($B_4$$>$0, $A_4$$<$0).  
The strong boxiness of the M31 bulge is evident not only by its Fourier decomposition (Fig.\ 3)
but by the difference between the galaxy isophotes and their corresponding 
best-fitting ellipses (Fig.\ 2).
The ELLIPSE algorithm also determines the position angle (PA) of the major axis of each isophote 
and its ellipticity ($e=1-b/a$); these are shown in Figure 3.
We discuss the full NIR M31 surface brightness profile derived 
from the 6x data, and its bulge/disk decomposition,  
elsewhere.

Based on trends in Figures 2--3, 
we discriminate four observational regions by semimajor axis radius within the inner $\sim$$3300$$''$ of M31
(see divisions in Fig.\ 3): 
%
%The Nucleus
%
(1) A moderately circular nucleus dominates the  central $<3$$''$.  The
nucleus was shown to be an independent feature of M31 (Light et al.\ 1974), 
so the characteristics within 3" do not bear on the the morphology of the bulge. 
In any case, because of the 1$''$ pixel sampling, the PA and Fourier decomposition here
have little meaning. 
%
%The Circular Bulge
%
(2) A near-circular bulge (3$''$ to 50$''$) 
resembling a classical bulge with low ellipticity and some
diskiness.  The primary 
structural characteristic of this radial zone is a 20$^{\circ}$ change in 
position angle.  This isophotal twisting is evident by the slight kinking 
of the innermost parts of the ``Maltese cross" pattern in the
bottom panel of Figure 2.  
%
%The Boxy Bulge
%
(3) The boxy bulge ($\sim50$$''$ to $\sim$700$''$) exhibits distinctly 
negative values of $B_4/B$ and a steadily increasing
ellipticity with radius (at constant PA). 
The strongest boxiness, at $\sim$$300$$''$, is evident by the width and
amplitude of the Maltese cross arms at this point.
%
%The Inner Disk
%
Finally, (4) a transition to the inner disk ($>700$$''$) is demarcated 
by the shift to disky Fourier components and by the lower 
surface brightness evident in Figure 1.  
Because of the lower surface brightness and partial coverage,
the PA and ellipticity are less reliably measured here.  
However, an obvious feature in this radial zone is the narrow bar
structure protruding 
from both ends of the boxy bulge, shifted to nearly the {\it disk} PA,
and extending to a radius of $\sim$1200$''$ (4 kpc) or more.

Several trends hold throughout the bulge/bar regions just described: 
(a) The three infrared bands closely track one another in morphology and relative
brightness, so that no color gradients are evident outside the nucleus
(this color uniformity is evident by the nearly uniform whiteness of the
central structures in the color composite in Fig.\ 1), 
(b) the center of the bulge isophotes for different NIR bands are 
coincident at the center of the optical disk,  
(c) the primary position angle of the
bulge, about $50^{\circ}$, is about 10$^{\circ}$ higher than the PA of the outer disk, and 
(d) the boxiness of M31's bulge rivals that seen in any galaxy.

\section{Discussion}

This new large scale infrared mapping of M31 vividly highlights the morphology of
its bulge, definitively confirms earlier suggestions from optical studies 
(\S1) of a boxy inner structure, and permits high $S/N$ measurements of bulge 
shape 
without interference from variable dust extinction (Fig. 3).  M31 clearly twins 
the Milky Way (Hauser et al.\ 1990, Weiland et al.\ 1994) --- as well as a large 
fraction (45\%, according to \citeauthor{lut2} \citeyear{lut2}) of other highly 
inclined disk galaxies --- in having a boxy inner structure that is made more 
evident by infrared imaging.  Indeed, $K$ band images by \citeauthor{shaw93} 
(\citeyear{shaw93}) and Bureau et al. (2006) of nearly 30 edge-on systems with a boxy/peanut bulge all 
display residual maps with Maltese cross patterns 
closely resembling that observed here in M31, and, when sufficiently face-on, 
isophotal twists.  
Boxy bulges are now recognized from $N$-body simulations to represent the 
vertical structure of bars viewed at high inclination (Combes et al.\ 1990; 
see review by Athanassoula 2005).
This is clearly demonstrated by the Milky Way, which appears to have a 
boxy bulge similar to the one we observe in M31 
(Binney, Gerhard \& Spergel 1997, Bissantz \& Gerhard 2002).  
%Despite this size differential, 
The M31 inner structure can also be reproduced by self-consistent $N$-body 
modeling of a barred galaxy, as long as a classical bulge is also present
(Athanassoula \& Beaton 2006, see inset to Fig.\ 1).  In this model, the bar 
creates and extends well past the 700$''$ boxy bulge, and may even account 
for the 50$'$ radius ``pseudo-ring" of M31.  The 6x images show clear 
evidence for this ``thin bar" extending to at least $\sim$1200$''$ (4 kpc).
From the presence of this feature, together with the successful modeling of 
the images presented here using a bar potential as well as 
the numerous other examples of models and 
data indicating the close connection between bars and boxy bulges, it may be 
safely concluded that M31 is a barred spiral galaxy.

%\section{Acknowledgements}

\vskip 0.15in
This publication makes use of data products from the Two Micron All Sky Survey, which is a joint project of the 
University of Massachusetts and the Infrared Processing and Analysis Center/California Institute of Technology, 
funded by the National Aeronautics and Space Administration and the National Science Foundation (NSF). 
We appreciate support by NSF grant AST-0307851 and a Space Interferometry Mission Key Project grant, 
NASA/JPL contract 1228235.  RLB was supported by a Harrison Undergraduate Research Award from the University of Virginia
Center for Undergraduate Research.  MFS acknowledges support from NASA/JPL contract 1234021. This work was also 
supported by the F.H. Levinson Fund of the Peninsular Community Foundation. 
 P.G.\ was supported by NSF grants AST-0307966 and AST-0507483 and
       NASA/STScI grants GO-10265.02 and GO-10134.02.

%We thank A. Bosma, M. Bureau, P. Patsis, G. Aronica and Ch. Skokos for stimulating discussions on boxy bulges and bars. EA thanks the INSU/CNRS, the region PACA and the University of Aix-Marseille I for funds to develop the computing facilities used for the simulations discussed in this paper.

\end{document}